\documentclass[aps,preprint,groupedaddress,showpacs,showkeys,floatfix]{revtex4-1}

\usepackage{graphicx}
\usepackage{dcolumn}
\usepackage{amsmath}
\usepackage{amssymb}
\usepackage{epstopdf}
\usepackage{rotating}
\usepackage{color}
\bibpunct{[}{]}{,}{n}{}{}

\newcommand{\bft}{\textit{Boltzmann Factor} tutorial}
\newcommand{\BF}{Boltzmann factor}

\newcommand\stat{\textit{Stat Mech}}
\newcommand{\eval}[1]{\textrm{\hspace{0.5mm}\rule[-3mm]{0.25mm}{8mm}}_{_{ \hspace{0.25mm} #1 }}\textrm{\hspace{-2.5mm}}}
\newcommand{\evaln}[1]{\textrm{\hspace{0.5mm}\rule[-3mm]{0.25mm}{8mm}}_{ \hspace{0.5mm} #1 }\textrm{\hspace{-1mm}}}

\begin{document}

\title{Student Understanding of Taylor Series Expansions in Statistical Mechanics}


\author{Trevor I. Smith}\address{Department of Physics \& Astronomy, Dickinson College, Carlisle, PA 17013}

\author{John R. Thompson}\address{Department of Physics \& Astronomy, University of Maine, Orono, ME 04469}\address{Maine Center for Research in STEM Education, University of Maine, Orono, ME 04469}

\author{Donald B. Mountcastle}\address{Department of Physics \& Astronomy, University of Maine, Orono, ME 04469}

\begin{abstract}
One goal of physics instruction is to have students learn to make physical meaning of specific mathematical ideas, concepts, and procedures in different physical settings. As part of research investigating student learning in statistical physics, we are developing curriculum materials that guide students through a derivation of the Boltzmann factor, using a Taylor series expansion of entropy. Using results from written surveys, classroom observations, and both individual think-aloud and teaching interviews, we present evidence that many students can recognize and interpret series expansions, but they often lack fluency with the Taylor series despite previous exposures in both calculus and physics courses. We present students' successes and failures both using and interpreting Taylor series expansions in a variety of contexts.
\end{abstract}

\maketitle

\section{Introduction}
As part of an ongoing collaborative project we have documented student difficulties in and related to thermal physics. We have also made efforts to address these difficulties by developing and implementing guided-inquiry worksheet activities (a.k.a.\ tutorials). One branch of this research has focused on student understanding of the mathematics used in thermal physics, with particular emphasis on their knowledge of integration, differentiation, and mixed second-order partial derivatives.\cite{Thompson2006,Bucy2007a,Pollock2007} In this paper we report on a study in which we investigated students' understanding of the Taylor series both in a generic mathematical context and in a specific physical context.

The Taylor series expansion of a function, $f(x)$, about a given value, $x=a$, is a power series in which each coefficient is related to a derivative of $f(x)$ with respect to $x$. The generic form of the Taylor series of $f(x)$ about the point $x=a$ is,
\begin{align}
f(x)&=f(a)+\frac{\mathrm{d}f}{\mathrm{d}x}\evaln{a}(x-a)+\frac{1}{2}\frac{\mathrm{d}^2f}{\mathrm{d}x^2}\evaln{a}(x-a)^2+\dots\label{Taylor_math}\\
&=\sum\limits_{n=0}^\infty\frac{1}{n!}\frac{\mathrm{d}^nf}{\mathrm{d}x^n}\evaln{a}(x-a)^n.\nonumber
\end{align}
The series is often truncated by choosing a finite upper limit for the summation based on the specific scenario in which the expansion is being implemented. 

The Taylor series expansion is used ubiquitously throughout physics to help solve problems in a tractable way.  It is the mathematical root of several well known formulas across physics, ranging from one-dimensional kinematics, 
\begin{equation}
x(t)=x_0+v_0(t-t_0)+\frac{1}{2}a(t-t_0)^2,\label{kineq}
\end{equation}
through electricity and magnetism and quantum mechanics.

Other common uses of Taylor series expansions include numerical computations, evaluations of definite integrals and/or indeterminate limits, and approximations.\cite{Boas1983} Approximations are particularly useful in physics at times when a solution in its exact form is unnecessary or too difficult to obtain; in situations where information is known about various derivatives of a function at a specific point, but nothing more is known about the function itself; or in situations in which one is investigating sufficiently small fluctuations about (or changes to) an average value. We are particularly interested in students' understanding of these uses in the context of expanding the entropy of a relatively large thermodynamic system as a function of the energy of a much smaller system with which it is in thermal equilibrium.

We have previously reported various student difficulties using the \BF\ in appropriate contexts as well as efforts made to address these difficulties.\cite{Smith2010} Our main effort has been the creation of a tutorial that leads students through a derivation of the \BF\ and the canonical partition function.\footnote{The derivation in the \bft\ closely resembles that found in many thermal physics textbooks, cf.\ Ref.\ \citealp{Baierlein1999}.} As part of the derivation, the students must write a Taylor series expansion of entropy as a function of energy. Before the first implementation of the \bft\ we anticipated that students might have difficulty generating this Taylor series. This led to the development of a pretest on the graphical interpretation of a Taylor series expansion as well as a pre-tutorial homework assignment for students to complete on their own and bring to class for use during the tutorial. 

We have two main findings from this work, one focused on student understanding
and one related to pedagogical strategy. First, data from written surveys and clinical interviews suggest that many students are familiar with the Taylor series but may not use it fluently in physical contexts. Second, results from teaching interviews and several years of tutorial implementation suggest that the pre-tutorial homework assignment provides students with the necessary opportunity to refresh their memory of what exactly a Taylor series is and how to use it in physical contexts.  

We begin with a brief overview of the research methods used for this study. We present results from survey data on students' interpretation of Taylor series expansions. We continue with data on students' use of Taylor series within the \bft\ (both during classroom implementations and interviews). Finally, we discuss data from interviews on student understanding of Taylor series within the general context of physics. We conclude with a summary of our findings and suggestions for future research opportunities.

\section{Research Methods}
Data for this study were gathered during three subsequent years in an upper-division undergraduate-level statistical mechanics course (\stat) at a public research university in the northeastern United States. The population under investigation was comprised primarily of junior or senior undergraduate physics majors and graduate students in physics. \stat\ contains no explicit instruction on Taylor series, but a prerequisite course in mathematics does. Data were collected using written surveys, teaching and clinical interviews, and classroom observations.

A written survey was administered before the \bft\ each year to assess students' ability to interpret a Taylor series expansion of a function, given a graph of that function (Taylor series pretest, or ``pretest'').\footnote{The Taylor series pretest was developed by Warren Christensen, based on a suggestion by Andrew Boudreaux.} These data provided insight into students' prior knowledge regarding Taylor series expansions. 

Data from classroom observations were gathered each year by videotaping students participating in the \bft. These data were used to examine how students used the Taylor series to progress through the derivation in the tutorial.

To gain more insight into individual students' understanding of the Taylor series, we conducted two sets of student interviews. The first set was conducted in the style of a teaching experiment, in which the interviewer acted as an instructor, offering progressively more assistance.\cite{Steffe2000} In these individual teaching interviews, which took place shortly after the Year 1 implementation of the \bft, four students were asked to complete portions of the tutorial (including a Taylor series generation) while thinking aloud. The interviewer offered guidance only when needed. The goal of the teaching interviews was to determine how well students could complete the tutorial activities and how much guidance was necessary. 

The interviews in the second set were conducted in the style of a clinical interview to assess students' knowledge and understanding of the Taylor series. The clinical interviews differed from the teaching interviews in that the interviewer was careful to minimize influence on student responses by interview prompts.\cite{Piaget1967} By studying students' ideas in the absence of teaching we gain a better sense of what students really understand, rather than what they can reflect back to an instructor. Clinical interviews were conducted with five students, several weeks after the Year 2 implementation of the \bft. 

\section{Student Interpretation of a Taylor Series Expansion}
\label{sec:taylor}
\begin{figure}[tb]
\includegraphics[totalheight=0.15\textheight]{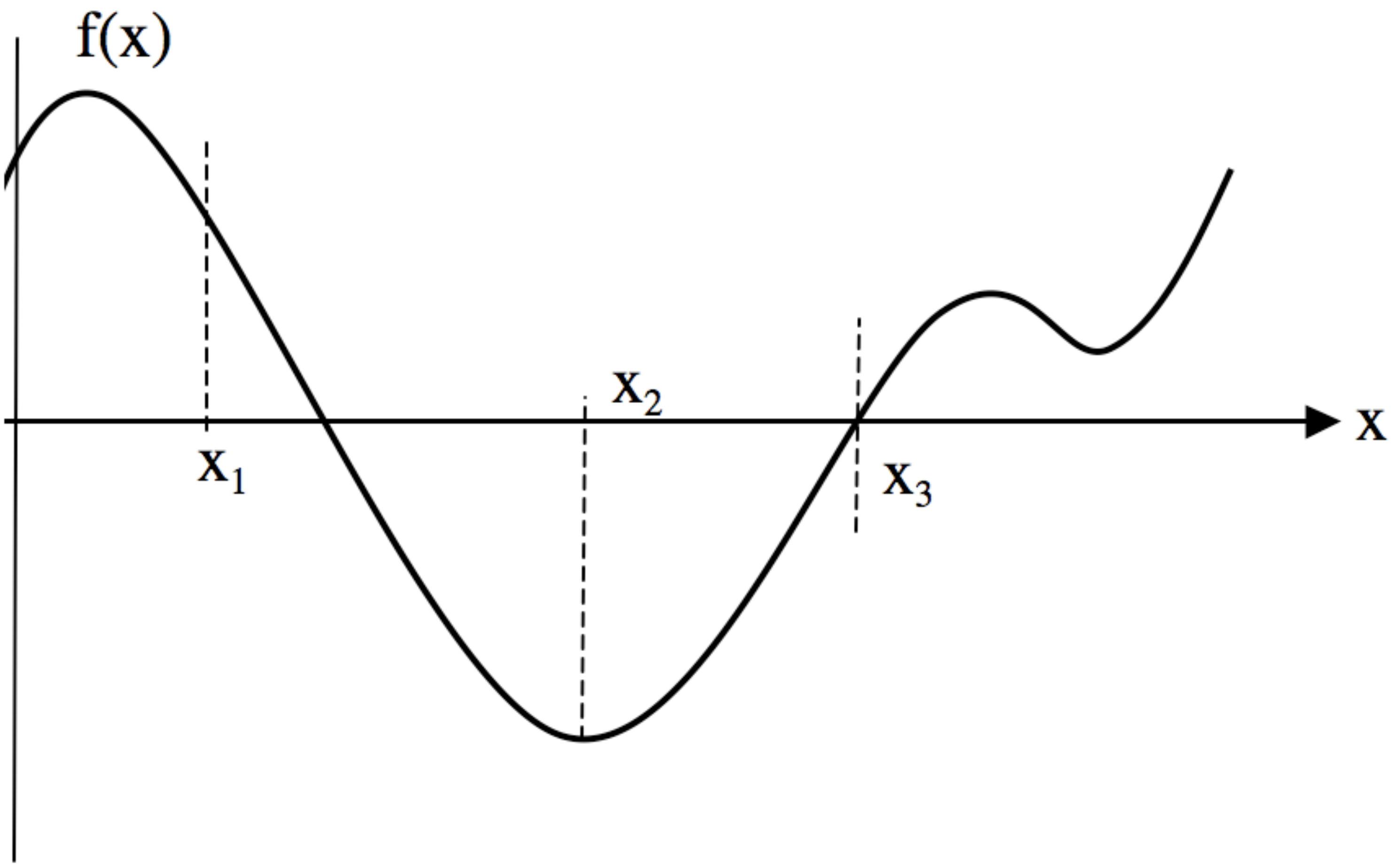}
\caption{Graph used in the Taylor series pretest.}
\label{taylor-pt}
\end{figure}

In the pretest, students interpret the terms of a Taylor series expansion based on a given graph of a function, $f(x)$, shown in Figure \ref{taylor-pt}. The truncated Taylor series expansion about the point $x\nobreak=x_1$ is given as
\begin{equation}
f(x)=a_1+b_1(x-x_1)+c_1(x-x_1)^2.\label{Taylor_PT}
\end{equation}
Students are asked to determine whether each of the quantities $a_1$, $b_1$, and $c_1$ are positive, negative, or zero and to explain their reasoning based on the graph with $x_1$ clearly marked. The same question is asked for the locations $x_2$ and $x_3$ on the graph. The correct answers require students to recognize that $a$ is the value of the function at the specified point, $b$ is the slope of the function (i.e., the first derivative), and $c$ is proportional to the concavity (corresponding to the second derivative) at that point. In the three years that this question was given before tutorial instruction was implemented, 14 out of 26 students correctly determined all the signs of the various quantities and gave appropriate reasons.\footnote{An additional 4 students in Year 3 gave responses that indicated that they probably understand Taylor series expansions. Three of these students gave 8 or 9 correct responses (out of 9) without any explanations; the other wrote the correct form for each coefficient but did not determine the signs of any of the coefficients. Without explanation, however, we do not consider these responses sufficient evidence that these students truly understand how to graphically interpret a Taylor series expansion.} These data suggest that about half of the students in \stat\ are familiar with the meanings of the various terms in the Taylor series.

\section{Student Use of Taylor Series in Context}
As mentioned above, teaching interviews were conducted with four students in Year 1 to determine their abilities to complete the tutorial tasks.\cite{Smith2010} In the tutorial, students are asked to complete the derivation of the \BF, which involves a small system in thermal contact with a large reservoir (e.g., Ref.\ \citealp{Baierlein1999}). The derivation requires students to generate a Taylor series expansion of the entropy of the reservoir ($S_{res}$) as a function of its energy ($E_{res}$). 
Interviewed students were presented with this physical situation and asked to write a Taylor series expansion of the entropy of the reservoir as a function of its energy about the (fixed) total energy ($E_{tot}=E_{res}+E_{sys}$): 
\begin{align}
S_{res}(E_{res})&=S_{res}(E_{tot})+\frac{\partial S_{res}}{\partial E_{res}}\eval{E_{_{tot}}}\textrm{\hspace{-2mm}}(E_{res}-E_{tot})+\ldots \nonumber\\
&=S_{res}(E_{tot})-\frac{1}{T}E_{sys}+O(E_{sys}^2),\label{Taylor_CR}
\end{align}
where the simplification in the second line comes from the considerations that (a) the system and the reservoir have the same constant temperature, (b) the energy of the system can be written as $E_{sys}=E_{tot}-E_{res}$, and (c) $E_{sys}\ll E_{res}$. The second line of Eq.\ \ref{Taylor_CR} is a valuable result, as it relates the entropy of the reservoir to the energy of the system, which is (typically) much easier to measure.

During the four teaching interviews, only one student (Joel\footnote{All student names are pseudonyms.}) succeeded in spontaneously generating a Taylor series expansion of reservoir entropy as a function of energy (as in Eq.\ \ref{Taylor_CR}); however, other portions of Joel's interview suggest that his success was more a result of memorizing the derivation of the \BF\ from the course textbook rather than evidence of thorough comprehension of the Taylor series.\cite{Smith2010,Smith2011} Two other students were able to generate the appropriate expansion only after they were given an expression for a Taylor series expansion of entropy as a function of energy about the value $E=E_0$,
\begin{align}
S(E)=S(E_0)+&\frac{\partial S}{\partial E}\eval{E_{_{0}}}(E-E_0)\label{Taylor_gen}\\
&+\frac{1}{2!}\frac{\partial^2 S}{\partial E^2}\eval{E_{_{0}}}(E-E_0)^2+\ldots\nonumber
\end{align}
The final student was also given this expression but was unable to connect it with the specific physical scenario without explicit instruction from the interviewer. 

These results indicate that student understanding of when and how to use a Taylor series expansion (a crucial part of the derivation used in the \bft) cannot be taken for granted. When combined with the data from the Taylor series pretest discussed above, these results indicate that students may be able to interpret and apply a Taylor series that is given to them, but not be able to generate an appropriate expansion in a novel context. These difficulties, however, were not completely unexpected. 

One instructional strategy that we incorporated into our curriculum development is the use of what we call ``pre-tutorial homework.''  We have seen that, at the upper division in particular, there is much detailed prerequisite knowledge --- both mathematical and physical --- that needs to be readily accessible in order for students to make appropriate progress through the tutorial in the allotted time. One way we have found to address this issue is to assign homework to be completed prior to the tutorial, in which students engage with the prerequisite knowledge at their leisure. 

We developed a two-question pre-tutorial homework assignment for the \bft.\footnote{The pre-tutorial homework was assigned after students had completed the \bft\ pretest (which included the Taylor series pretest question and additional questions, which are not presented in this paper).} In the first question, students are asked to write a Taylor series expansion of entropy as a function of energy (including no more than five terms) about the value $E=E_0$ (see Eq.\ \ref{Taylor_gen}). This gives students the opportunity to look up the generic form of the Taylor series (cf.\ Eq.\ \ref{Taylor_math}) and apply it to the expansion of entropy as a function of energy (cf.~Eq.~\ref{Taylor_gen}). They would then have that formula with them in class to apply to the specific physical situation presented in the tutorial, thus deriving Eq.\ \ref{Taylor_CR}. 



\begin{figure}[tb]
\includegraphics[totalheight=0.2\textheight]{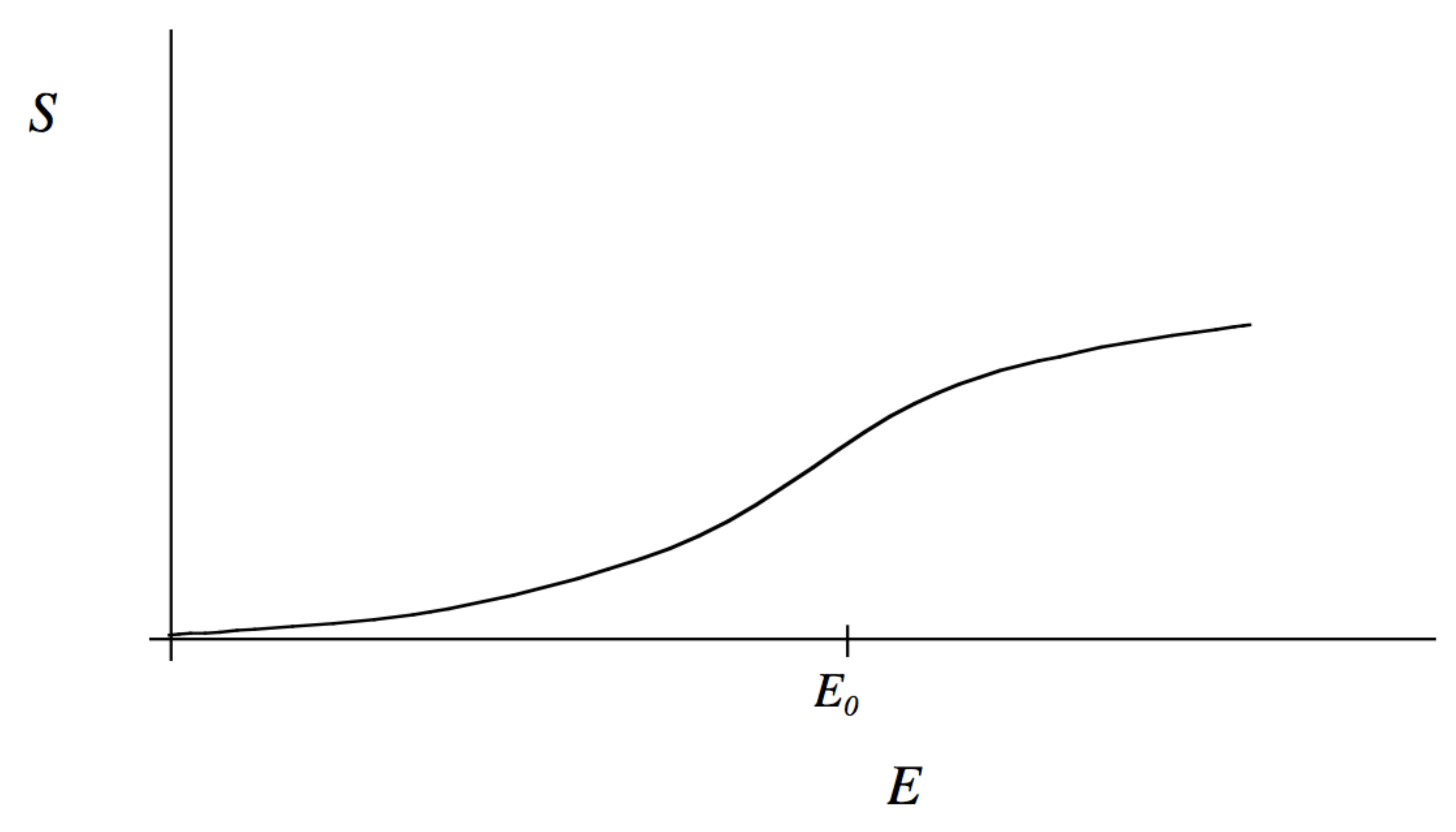}
\caption{Graph used in the pre-tutorial homework assignment. Students were asked to give a graphical interpretation of each of the terms in their Taylor series expansion of entropy as a function of energy about $E=E_0$ (see Eq.\ \ref{Taylor_gen}).}
\label{Taylor-prehw}
\end{figure}

In the second homework question, students are asked to give an interpretation of how each of the terms in the Taylor series relates to a given graph of $S$~vs.~$E$ (see Figure \ref{Taylor-prehw}). This question encourages students to think about the meaning of the terms in the Taylor series rather than merely copying down abstract symbols.

One unexpected difficulty observed during the tutorial session in Year 2 is that two students did not correctly construct the Taylor series expansion asked for in the pre-tutorial homework assignment. Instead of constructing the appropriate expansion as seen in Eq.\ \ref{Taylor_gen}, they used the terms ``$S_1$'' and ``$S_2$'' in place of the ``$(E-E_0)$'' and ``$(E\nobreak-\nobreak E_0)^2\,$'' terms, respectively (i.e., $S=S_0+S_1S^{\prime}+\frac{1}{2}S_2S^{\prime\prime}+\dots$, where $S_0$, $S_1$, and $S_2$ were said to be constants). This expansion not only made it impossible for the students to obtain an expression for entropy in terms of energy, but also prohibited them from obtaining a dimensionally accurate expression for entropy at all. These students did, however, recognize that their expression lacked an energy term, and once an instructor intervened to discuss the appropriate form of the Taylor series with them, they were able to use it correctly to complete the derivation of the \BF. This is further evidence that the successful completion of the task given on the pre-tutorial homework assignment is crucial to student success with the tutorial, yet the assignment itself is not trivial, as some students may fail to complete the task.

During Year 2 the \bft\ was also implemented in an upper-division undergraduate-level thermal physics course at a comprehensive public university in the western United States. The instructor of this course reported that many of his students had great difficulty using the Taylor series expansion in the tutorial context even after having completed the pre-tutorial homework. In an effort to help the students the instructor split the \bft\ into two class periods and assigned specific study of the Taylor series between the two periods. He reported that a short lecture on the use of Taylor series expansion was necessary at the beginning of the second tutorial period to allow students to successfully complete the tutorial. This report provides further evidence that the use of the Taylor series in this context is not trivial and suggests that student difficulties in this area are not localized to our student population.

\section{Further Investigation into Student Understanding of the Taylor Series}
After tutorial implementation in Year 2, clinical interviews were conducted with five students, four of whom had participated in the \bft\ in class. One of the goals of the interviews was to determine how familiar the students were with Taylor series expansions, including when they are applicable and how they are used. 

All students interviewed had a reasonable understanding of situations in which the Taylor series is an appropriate tool. All students spontaneously used terms like ``approximation'' and ``estimation'' when describing how to use a Taylor series expansion, and all students were able to list one or more specific areas of physics in which Taylor series expansions are useful. One interesting aspect of the interviews is that all students at some point during the interview \textit{spontaneously} referred to the kinematic equation (Eq.\ \ref{kineq}) as an example of the Taylor series. This had been described during \stat\ lecture as an example of a Taylor series expansion with which everyone would be familiar (even if they had never thought of it as a Taylor series). Their acknowledgment of that kinematic equation as a Taylor series seemed to influence their responses to various interview prompts.

One of the main questions the interviewees were asked about the Taylor series was how they knew when to truncate the series. A common response involved knowledge about the functional form of any higher order derivatives; i.e., if one of the derivatives is constant, then all higher derivatives will be identically zero. One student (Malcolm, a graduate student in physics) used this reasoning to justify why the kinematic equation has only three terms: ``Usually acceleration's constant, so we don't have a jerk. If we had a jerk running around messing things up, we'd need more terms.'' When prompted about situations in which no information was known about the derivatives, however, Malcolm said that he would use different ``rules of thumb'' depending on the application. If only a ``ballpark'' estimate was needed, for example, only one or two terms would be necessary, but he indicated more terms would be needed as desired precision increased (e.g., to ``16 digits''). Malcolm also expressed the idea that looking very close to the value about which he was expanding would require fewer terms than if he were trying to extrapolate too far from the expansion point. Finally, Malcolm stated that he would examine the deviation between the Taylor series expansion and any experimental data available and keep enough terms to have a reasonable fit (although he did not specify how close he would require the expansion to match experimental data). 

The relation to experimental data was echoed by Jayne (another graduate student in physics) who initially had trouble articulating a good rationale for truncating a series but eventually referred to different needs for different experimental tasks. Jayne cited a threshold for truncation of three or four orders of magnitude, i.e., terms that are 3--4 orders of magnitude smaller than the linear term are not necessary.

Two undergraduate physics majors who were interviewed together (Paul, who participated in the \bft, and Jonah, who did not) also cited constant acceleration as the reason why the kinematic equation only has three terms and knowledge of constant derivatives as the primary reason to truncate a Taylor series. After several prods and questions about series truncation they started using ``estimation'' language to discuss the possibility of starting with a ``ballpark'' estimate and keeping terms until the results were close enough (using a guess-and-check type of method). Paul also argued that the purpose of a Taylor series is to estimate something that is more complex and that the first few terms should be the most significant while the higher-order terms die out.

All students interviewed were able to list some areas of physics in which the Taylor series might be useful other than the kinematic equation (examples in quantum mechanics, solid state physics, statistical mechanics, etc.), but no one elaborated on exactly \textit{how} a Taylor series would be useful in these various situations. Malcolm came closest by citing the use of Taylor series to approximate a potential in quantum mechanics as a harmonic potential, a task he implied he had completed in the past. It is still unclear, however, what (if anything) would motivate these students to spontaneously use a Taylor series expansion in a given physical situation. We do not have evidence that they are able to generalize their knowledge to state the conditions under which a Taylor series is appropriate, and when to terminate one. It seems as though their past experience has been based on various instructors and texts specifying both when a Taylor series is appropriate and how many terms are necessary. 

\section{Conclusions and Implications for Future Work}
Our studies on student understanding of a Taylor series expansion as it applies to thermal physics have provided mixed results. Data from the Taylor series pretest (see Figure \ref{taylor-pt}) indicate that many students were able to interpret a Taylor series of a function given the graph of that function. Results from interviews and classroom observations, however, indicate that students struggled with generating a Taylor series expansion in a physical context (e.g., with entropy and energy). Once provided with a generic Taylor series using physical quantities, most students were able to apply it to a specific situation, but this was usually not a trivial task for them (as seen during the teaching interviews and classroom observations). 

These results underscore the need for the pre-tutorial homework assignment in which students are asked to generate the Taylor expansion in Eq.\ \ref{Taylor_gen}. We have found this pre-tutorial homework strategy to be worthwhile, even necessary, for implementing tutorials in upper-division thermal physics courses. This marks a distinct difference from typical tutorial implementation in introductory courses, as far more prerequisite knowledge is both required and assumed at the upper division, including a robust understanding of concepts in both physics and mathematics.

Results from clinical interviews on student understanding of the applicability of Taylor series expansions show that many students recognized that the Taylor series is a relevant mathematical tool in various areas of physics, but they often lacked a sense of when its use is appropriate. Students also did not have rigorous criteria for determining how many terms should be kept (except when one of the derivatives is a constant, resulting in all higher derivatives being identically zero).

It is worthwhile to point out that application of a Taylor series expansion to specific physical scenarios, especially when the quantities involved are entropy and energy, is in our opinion a very sophisticated and complex process. It involves not only recall of the mathematical expansion, but an understanding of the mathematical meaning of each term, a physical interpretation of each of those terms as related to the scenario at hand, and judgment of the appropriate conditions to apply for series termination. This also assumes that a student recognizes that a Taylor series expansion could  make the problem tractable.  When the physical quantities are abstract and known to be difficult to understand conceptually, the task is that much more challenging for students.  

We suggest several additional research questions based on our results: Under what conditions might students choose (or are able) to use a Taylor series expansion without instructor intervention? What aspects of a physical scenario should be highlighted to encourage its use? Are some physical quantities easier for students to use in a Taylor series expansion? That is, perhaps entropy and energy are too abstract for our students to use with sophisticated mathematical tools like Taylor series expansion. Of particular interest would be an examination of expert physicists' spontaneous use of the Taylor series. By learning when expert physicists choose to use Taylor series and how they make that decision, one could (in principle) design an instructional sequence that could enhance student understanding of physics and useful mathematical tools within many different courses typically taught in the undergraduate sequence. This would be an excellent stepping stone for undergraduate physics majors on their way to expertise. 

\begin{acknowledgments}
We thank Warren Christensen and Andrew Boudreaux for the inspiration for and development of the Taylor series pretest. We are grateful to members of the University of Maine Physics Education Research Laboratory, especially Warren Christensen, and our other colleagues, Michael Loverude and David Meltzer, for their continued collaboration and feedback on this work.  This material is based upon work supported by the National Science Foundation under Grant No.\ DUE-0817282.
\end{acknowledgments}

\bibliography{MCWTIS}
\bibliographystyle{apsrev4-1}   

\end{document}